\documentclass{svmult}

\usepackage{mathptmx}       % selects Times Roman as basic font
\usepackage{helvet}         % selects Helvetica as sans-serif font
\usepackage{courier}        % selects Courier as typewriter font
\usepackage{type1cm}        % activate if the above 3 fonts are
\usepackage{makeidx}         % allows index generation
\usepackage{graphicx}        % standard LaTeX graphics tool
\usepackage{multicol}        % used for the two-column index
\usepackage[bottom]{footmisc}% places footnotes at page bottom

\usepackage{natbib}
\bibpunct{(}{)}{;}{a}{}{,}

\newcommand{\msun}{{M_\odot}}

\newcommand{\farcs}{\mbox{\ensuremath{.\!\!^{\prime\prime}}}}%  % fractional arcsecond symbol: 0.''0

\makeindex             % used for the subject index
\begin{document}

\title*{The Formation of Very Massive Stars}
% Use \titlerunning{Short Title} for an abbreviated version of
% your contribution title if the original one is too long
\author{Mark R.~Krumholz}
% Use \authorrunning{Short Title} for an abbreviated version of
% your contribution title if the original one is too long
\institute{Mark R.~Krumholz \at Department of Astronomy \& Astrophysics, University of California, Santa Cruz, CA 95064 USA, \email{mkrumhol@ucsc.edu}}
%
% Use the package "url.sty" to avoid
% problems with special characters
% used in your e-mail or web address
%
\maketitle

\abstract*{In this chapter I review theoretical models for the formation of very massive stars. After a brief overview of some relevant observations, I spend the bulk of the chapter describing two possible routes to the formation of very massive stars: formation via gas accretion, and formation via collisions between smaller stars. For direct accretion, I discuss the problems of how interstellar gas may be prevented from fragmenting so that it is available for incorporation into a single very massive star, and I discuss the problems presented for massive star formation by feedback in the form of radiation pressure, photoionization, and stellar winds. For collision, I discuss several mechanisms by which stars might be induced to collide, and I discuss what sorts of environments are required to enable each of these mechanisms to function. I then compare the direct accretion and collision scenarios, and discuss possible observational signatures that could be used to distinguish between them. Finally, I come to the question of whether the process of star formation sets any upper limits on the masses of stars that can form.}

\abstract{In this chapter I review theoretical models for the formation of very massive stars. After a brief overview of some relevant observations, I spend the bulk of the chapter describing two possible routes to the formation of very massive stars: formation via gas accretion, and formation via collisions between smaller stars. For direct accretion, I discuss the problems of how interstellar gas may be prevented from fragmenting so that it is available for incorporation into a single very massive star, and I discuss the problems presented for massive star formation by feedback in the form of radiation pressure, photoionization, and stellar winds. For collision, I discuss several mechanisms by which stars might be induced to collide, and I discuss what sorts of environments are required to enable each of these mechanisms to function. I then compare the direct accretion and collision scenarios, and discuss possible observational signatures that could be used to distinguish between them. Finally, I come to the question of whether the process of star formation sets any upper limits on the masses of stars that can form.}

\section{Introduction}
\label{sec:intro}

The mechanism by which the most massive stars form, and whether there is an upper limit to the mass of star that this mechanism can produce, has been a problem in astrophysics since the pioneering works of \citet{larson71a} and \citet{kahn74a}. These authors focused on the physical mechanisms that might inhibit accretion onto stars as they accreted interstellar matter, and we will return to this topic below. However, a more modern approach to the problem of very massive stars requires placing them in the context of a broader theory of the stellar initial mass function (IMF).

The IMF is characterized by a peak in the range $0.1-1$ $\msun$, and a powerlaw tail at higher masses of the form $dn/d\ln m\propto m^\Gamma$ with $\Gamma\approx -1.35$ \citep[and references therein]{bastian10a}. However, the mass to which this simple powerlaw extends is not very well-determined. It is not possible to measure the IMF for field stars to very high masses due to uncertainties in star formation histories and the limited number of very massive stars available in the field. Measurements of the high-mass end of the IMF in young clusters must target very massive systems in order to achieve strong statistical significance, and such clusters are rare and thus distant. This creates significant problems with confusion. The limited studies that are available suggest that the a powerlaw with $\Gamma \approx -1.35$ remains a reasonable description of the IMF out to masses of $\sim 100$ $\msun$ or more \citep[e.g.][]{massey98b, kim06b, espinoza09a}. However, it is by no means implausible that there are hidden features lurking in the IMF at the highest masses. Indeed some analyses of the IMF have claimed to detect an upper cutoff (see the Chapter by F.~Martins in this volume for a critical review).

This observational question of whether the most massive stars are simply the ``tip of the iceberg" of the normal IMF, or whether they represent a fundamentally distinct population, animates the theoretical question about how such stars form. The two dominant models for how massive stars form are formation by accretion of interstellar material, i.e.~the same mechanism by which stars of low mass form, and formation by collisions between lower mass stars, which would represent a very different formation mechanism from the bulk of the stellar population.\footnote{Mergers between two members of a tight binary as a result of the growth of stellar radii during main sequence or post-main sequence evolution, or as a result of secular interactions in hierarchical triples, is a third possible mechanism by which massive stars can and probably do gain mass \citep{sana12a, de-mink13a, moeckel13a}. However, I do not discuss this possibility further, because it provides at most a factor of two increase in stellar mass.} In the remainder of this Chapter, I review each of these models in turn (Sections \ref{sec:accretion} and \ref{sec:collision}), pointing out its strengths, weaknesses, and areas of incompleteness. I then discuss the observable predictions made by each of these models, and which might be used to discriminate between them (Section \ref{sec:discrimination}). Finally, I summarize and return to the question first raised by \citet{larson71a} and \citet{kahn74a}: is there an upper mass limit for star formation, and if so, why (Section \ref{sec:masslimit})?

\section{The Formation of Very Massive Stars by Accretion}
\label{sec:accretion}

The great majority of stars form via the collapse of cold, gravitationally-unstable, molecular gas, and the subsequent accretion of cold gas onto the protostellar seeds that the collapse produces \citep[and references therein]{mckee07a}. There are numerous competing models for the origin of the observed $\Gamma\approx -1.35$ slope \citep[e.g.][]{bonnell01b, padoan02a, padoan07a, hennebelle08b, hennebelle09a, hennebelle13a, krumholz11c, krumholz12b, hopkins12d}, but in essentially all of these models, the massive end of this tail is populated by stars forming in rare, high-density regions that provide at least the potential for large mass reservoirs to be accreted onto protostellar seeds at high rates. Some but not all of these models identify the regions that give rise to massive stars with observed ``cores": compact ($\sim 0.01$ pc), dense ($>10^5$ molecules cm$^{-3}$) regions of gas, the largest of which can have masses large enough to form very massive stars \citep[e.g,.][]{beuther04b, beuther05b, bontemps10a}. None of these models predict that there is an upper limit to the masses of either cores or stars, and there is no observational evidence of a truncation either. Thus, it would seem that there is no barrier in terms of mass supply to the formation of very massive stars via the same accretion processes that give rise to the remainder of the IMF. However, the fact that there is a large supply of mass available does not guarantee that it can actually accrete onto a single object and thereby produce a very massive star. There are four major challenges to getting the available interstellar mass into a star, which we discuss below: fragmentation, radiation pressure, photoionization, and stellar winds. I discuss each of these challenges in turn in the remainder of this Section.

\subsection{Fragmentation}

The first challenge, fragmentation, can be stated very simply. When gravitationally-unstable media collapse, they tend to produce objects with a characteristic mass comparable to the Jeans mass,
\begin{equation}
\label{eq:jeansmass}
M_J = \frac{\pi}{6}\frac{c_s^3}{\sqrt{G^3\rho}} = 0.5 \left(\frac{T}{10\mbox{ K}}\right)^{3/2} \left(\frac{n}{10^4\mbox{ cm}^{-3}}\right)^{1/2} M_\odot,
\end{equation}
where $c_s$ is the sound speed, $\rho$ is the gas density, $T$ is the gas temperature, and $n$ is the gas number density. The temperature and density values to which I have scaled in the above equation are typical values in star-forming regions. Clearly, a massive star is an object whose mass is far larger than the Jeans mass of the interstellar gas from which it is forming. Why, then, does this gas not fragment into numerous small stars rather than forming a single large one? Indeed, hydrodynamic simulations of the collapse of compact, massive regions such as the observed massive cores show that they generally fail to produce massive stars \citep{dobbs05a}, and larger-scale simulations of star cluster formation appear to produce mass functions that are better described by truncated powerlaws than pure powerlaws \citep[e.g.][]{maschberger10a}, and where the formation of the most massive stars is limited by ``fragmentation-induced starvation" \citep{peters10b, girichidis12a}.

While these results might seem to present a serious challenge to the idea that massive stars form by accretion, they are mostly based on simulations that include no physics other than hydrodynamics and gravity. More recent simulations including a wider range of physical processes suggest that the fragmentation problem is much less severe than was once believed. Fragmentation is reduced by two primary effects: radiation feedback and magnetic fields.

Radiation feedback works to reduce fragmentation by heating the gas, raising its pressure and thus its Jeans mass (cf.~equation \ref{eq:jeansmass}). Although massive stars can of course produce a tremendous amount of heating, the more important effect from the standpoint of suppressing fragmentation is the early feedback provided by low mass stars, whose luminosities are dominated by accretion rather than internal energy generation. \citet{krumholz06b} first pointed out the importance of this effect, showing that even a $1$ $\msun$ star accreting at the relatively high rates expected in the dense regions where massive stars form could radiate strongly enough to raise the gas temperature by a factor of a few at distances of $\approx 1000$ AU. Since the minimum fragment mass is roughly the Jeans mass, and this varies as temperature to the $3/2$ power (equation \ref{eq:jeansmass}), this effect raises the minimum mass required for gas to fragment by a factor of $\approx 10$. Subsequent radiation-hydrodynamic simulations by a number of authors \citep{krumholz07a, krumholz10a, krumholz11c, bate09a, bate12a, offner09a} have confirmed that radiation feedback dramatically suppresses fragmentation compared to the results obtained in purely hydrodynamic models. \citet{krumholz08a} argue that this effect will efficiently suppress fragmentation in regions of high column density, allowing massive stars to form without their masses being limited by fragmentation. In contrast, \citet{peters10b} find that fragmentation limits the growth of massive stars even when heating by direct stellar photons is included, but their simulations do not include the dust-reprocessed radiation field that is likely more important for regulating fragmentation, and are limited to regions of much lower density than the typical environment of massive star formation.

Magnetic fields limit fragmentation in two ways. First, they remove angular momentum. In a collapsing cloud, the densest regions collapse fastest, and as the gas falls inward it attempts to rotate faster and faster in order to conserve angular momentum. When the collapsing gas is threaded by a magnetic field, however, the resulting differential rotation between inner collapsing regions and outer ones that have not yet begun to collapse twists the magnetic field lines. The twist produces a magnetic tension force that transfer angular momentum from the inner to the outer regions, a process known as magnetic braking. Formally, for an axisymmetric flow, one can show \citep[e.g.,][]{stahler05a} that the time rate of change of the angular momentum of a fluid element at a distance $\varpi$ from the rotation axis due to magnetic forces is given by
\begin{equation}
\frac{\partial J}{\partial t} = \frac{1}{4\pi} \left[B_\varpi \frac{\partial}{\partial \varpi} (\varpi B_\phi) + \varpi B_z \frac{\partial}{\partial z} B_\phi\right]
\end{equation}
where $\mathbf{B}$ is the magnetic field, and we have used cylindrical coordinates such that the components of $\mathbf{B}$ are $(B_\varpi, B_\phi, B_z)$. Thus in general if the toroidal ($\phi$) component of the magnetic field varies with either radial or vertical position, and the field also has a poloidal ($\varpi$ or $z$) component, there will be a magnetic torque that alters the angular momentum of the gas. For the types of magnetic field configurations produced by collapse, the net effect is to transport angular momentum outward. This process inhibits the formation of rotationally-flattened structures (e.g.~accretion disks). This is significant from the standpoint of fragmentation, because rotational flattening raises the density of the gas as it approaches the star, and dense, rotationally-flattened structures are vulnerable to the Toomre instability (see below), in which the self-gravity of a flattened rotation structure overcomes support from thermal pressure and angular momentum, leading to fragmentation and collapse.

Second, magnetic fields provide extra pressure support that prevents regions from collapsing unless their magnetic flux to mass ratios are below a critical value
\begin{equation}
\left(\frac{\Phi}{M}\right)_{\rm crit} = (4\pi^2 G)^{1/2}.
\end{equation}
Regions with masses small enough such that $\Phi/M < (\Phi/M)_{\rm crit}$ are said to be magnetically sub-critical, meaning that they do not have enough mass to overcome magnetic pressure support and collapse. Observations indicate that star-forming cores, over a wide range of size and density scales, tend to have flux to mass ratios that are roughly uniformly distributed from 0 up to $(\Phi/M)_{\rm crit}$ \citep[and references therein]{crutcher12a}. Thus the median core is magnetically supercritical, and is able to collapse despite magnetic support. However, gravity overcomes magnetic support only by a factor of $\sim 2$. If the flux to mass ratio is at all non-uniform, this implies that there may be significant amounts of mass contained in regions that are too magnetized to collapse. Simulations of massive protostellar cores by \citet{hennebelle11a} find that, for realistic levels of magnetization, the number of fragments is reduced by a factor of $\sim 2$ compared to a purely hydrodynamic calculation.

\begin{figure}[t]
\sidecaption
\centerline{\includegraphics[scale=0.8]{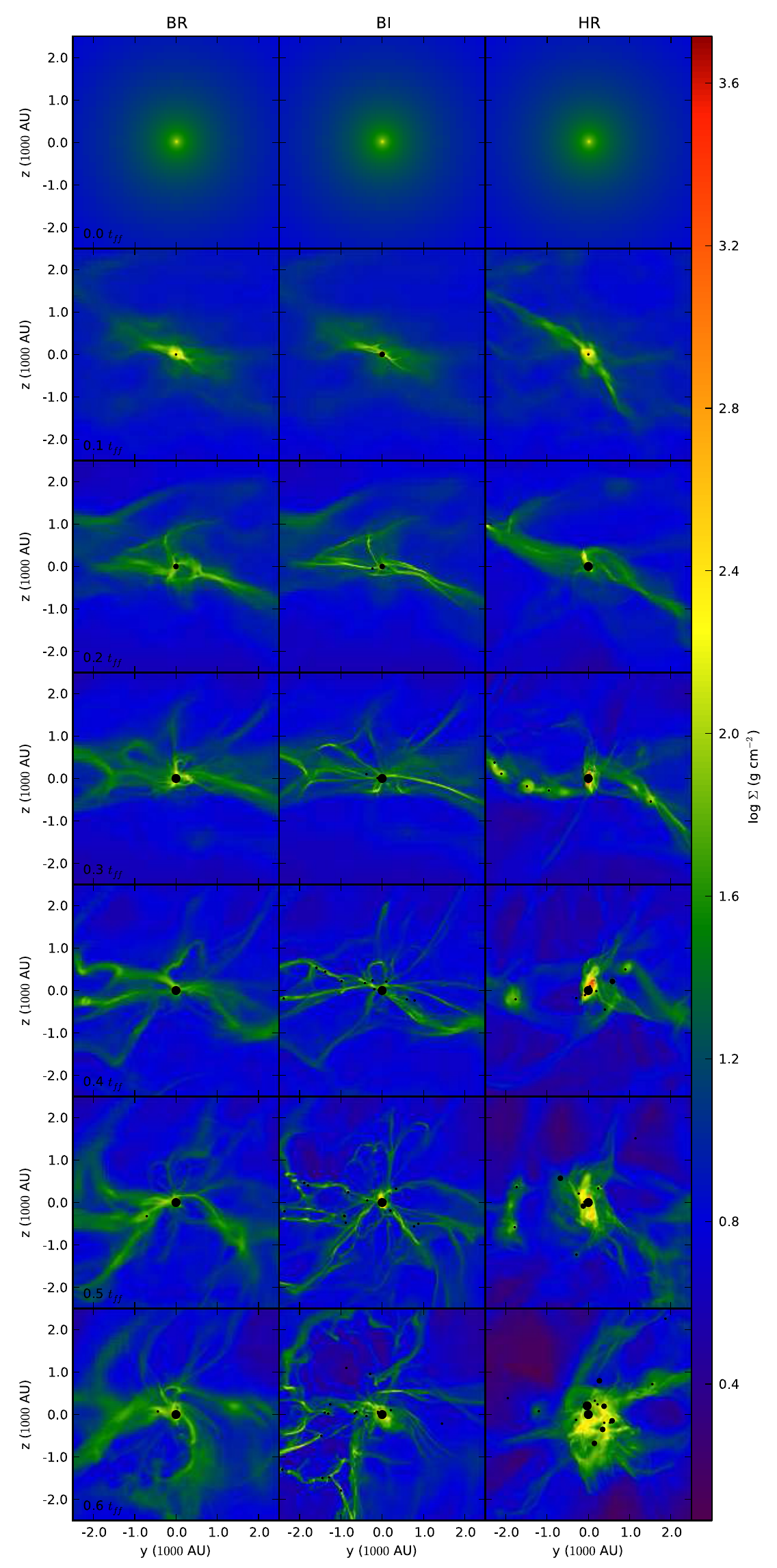}}
\caption{
Column densities from simulations of the collapse of massive protostellar cores \citep{myers13a}. The left column (BR) shows simulations including both magnetic fields and radiative feedback. The middle column (MI) uses magnetic fields but no radiation, while the right column (HR) uses radiation but has not magnetic field. Rows show snapshots at uniformly-spaced times, from the initial state to 0.6 core free-fall times. The region shown is the central 5000 AU around the most massive star. Colors show column density, and black circles show stars, with the size of the circle indicating the stellar mass. The initial magnetic field is oriented vertically in this projection. See \citet{myers13a} for more details.
\label{fig:myers13}
}
\end{figure}

More recently, \citet{commercon11c} and \citet{myers13a} have studied the collapse of massive cores using both radiative feedback and magnetic fields, and the effects amplify one another. At early times, the extra magnetic braking provided by magnetic fields removes angular momentum and channels material to the center faster. This tends to raise the accretion rate and thus the luminosity, making radiative heating more effective. Moreover, radiative and magnetic suppression of fragmentation are complementary in that they operate in different regions. Radiation suppresses fragmentation within $\approx 1000$ AU of a forming star, as found by \citet{krumholz06b} and subsequent radiation-hydrodynamic simulations, but becomes ineffective at larger radii. However, the regions more than $\approx 1000$ AU from a forming star are precisely those that are mostly likely to be magnetically sub-critical, and thus magnetic fields are able to suppress fragmentation in these regions. Because each mechanism operates where the other is weakest, the combination of the two reduces fragmentation much more efficiently than one might naively guess. Figure \ref{fig:myers13} shows an illustration of this effect: a simulation with magnetic fields and radiation shows almost no fragmentation, while ones with either alone both experience some fragmentation, though still less than in a purely hydrodynamic case. Based on these simulations, \citet{myers13a} conclude that compact, dense regions such as the observed massive cores are likely to form single star systems, rather than fragment strongly.

While this would seem to settle the question of whether fragmentation might limit stellar masses, it is worth noting that there is one final possible fragmentation mechanism that has not yet been evaluated via simulations. \citet{kratter06a} point out that the disks around massive stars are likely to be gravitationally-unstable. Gravitational stability for a pressure-supported disk can be characterized by the \citet{toomre64a} parameter,
\begin{equation}
Q = \frac{\kappa_{\rm ep} c_s}{\pi G \Sigma},
\end{equation}
where $\kappa_{\rm ep}$ is the epicyclic frequency (equal to the angular frequency of the orbit for a Keplerian disk), $c_s$ is the gas sound speed, and $\Sigma$ is the gas surface density. Values of $Q<1$ indicate instability of the disk to axisymmetric perturbations, and non-axisymmetric perturbations begin to appear at $Q\approx 1-2$. Depending on the properties of the disk, these instabilities can run away and cause the disk to fragment into point masses. For a steady disk with dimensionless \citet{shakura73a} viscosity $\alpha$, the accretion rate through the disk is \citep[e.g.,][]{kratter10a}
\begin{equation}
\dot{M} = \frac{3\alpha c_s^3}{GQ} = 1.5\times 10^{-4} \frac{\alpha}{Q}\left(\frac{T}{100\mbox{ K}}\right)^{3/2}\,M_\odot\mbox{ yr}^{-1},
\end{equation}
where the numerical evaluation for the sound speed uses $c_s=\sqrt{k_B T/\mu}$ and the mean particle mass $\mu=3.9\times 10^{-24}$ g, appropriate for fully molecular gas of standard cosmic composition. Local instabilities such as the magnetorotational instability in the disk cannot produce $\alpha > 1$, and the disk cannot be gravitationally stable if $Q<1$, so the accretion rate through a gravitationally-stable disk where angular momentum is transported primarily by local instabilities is strictly limited from above. Accretion rates of $10^{-4}$ $\msun$ yr$^{-1}$ in such disks are possible only if the temperature is $\approx 100$ K.

This means that there is a race between stellar heating and accretion. Forming a very massive star via accretion in a time less than its main sequence lifetime of a few Myr requires extremely high accretion rates -- $\sim 10^{-3}$ $\msun$ yr$^{-1}$ for a $>100$ $M_\odot$ star. However, such high accretion rates tend to be more than a disk can process without going unstable and fragmenting, unless radiation from the central star can heat the disk up, allowing it to transport mass more quickly while remaining stable. However, this process of heating to allow more mass through has a limit: once the temperature required to stabilize the disk exceeds the dust sublimation temperature, it will not be easy to heat the disk further, and this may result in an instability so violent that the disk fragments entirely, halting further accretion. \citet{kratter06a} estimate that this could limit stellar masses of $\sim 120$ $M_\odot$. Simulations thus far have not probed this possibility, as no 3D simulations have reached such high stellar masses. However, we caution that \citeauthor{kratter06a}'s scenario did not consider the effects of magnetic fields, which limit the disk radius and help stabilize it against fragmentation, or the effects of molecular opacity in the gas, which can provide coupling to the stellar radiation field and a means to heat the disk at temperatures above the dust sublimation temperature \citep{kuiper13a}.

\subsection{Radiation Pressure}

The second potential difficulty in forming massive stars via accretion is the radiation pressure problem, first pointed out by \citet{larson71a} and \citet{kahn74a}. The problem can be understood very simply: the inward gravitational force per unit mass exerted by a star of mass $M$ and luminosity $L$ on circumstellar material with specific opacity $\kappa$ located at a distance $r$ is $f_{\rm grav} = G M/r^2$, while the outward radiative force $f_{\rm rad} = \kappa L/(4\pi r^2 c)$. Since the radial dependence is the same, the net force will be inward only if
\begin{equation}
\frac{L}{M} < \frac{4\pi G c}{\kappa} = 2500\left(\frac{\kappa}{5\mbox{ cm}^2\mbox{ g}^{-1}}\right)^{-1} \,\frac{L_\odot}{M_\odot}.
\end{equation}
All stars above $\sim 20$ $\msun$ have $L/M > 2500$ $L_\odot/M_\odot$, so the question naturally arises: why doesn't radiation pressure expel circumstellar material and prevent stars from growing to masses substantially larger than $\sim 20$ $\msun$?

The choice of opacity $\kappa$ to use in evaluating this limit is somewhat subtle, because the dominant opacity source for circumstellar material will be dust that is mixed with the gas, which provides a highly non-gray opacity that will vary with position as starlight passes through the dust and is reprocessed by absorption and re-emission. Thus there is no single value of $\kappa$ that can be used in the equation above, and for an accurate result one must first compute the radiation field that results from the interaction of stellar photons with circumstellar dust, and then ask how the resulting radiation force compares to gravity. Nonetheless, detailed one-dimensional calculations by \citet{wolfire86a, wolfire87a}, \citet{preibisch95a}, and \citet{suttner99a}, including effects such as grain growth and drift relative to the gas, nonetheless confirm that radiation pressure is sufficient to halt accretion onto massive protostars at masses of $\sim 20$ $\msun$ for Milky Way dust abundances.

However, spherical symmetry is likely to be a very poor assumption for this problem, and a number of authors point out that relaxing it might reduce or eliminate the radiation pressure problem. The central idea behind these models is that the optically thick circumstellar matter around a rapidly-accreting protostar is capable of reshaping the radiation field emitted by the star, and making it non-spherical. If the radiation can be beamed, then the radiation force can be weaker than gravity over a significant solid angle even if the mean radiation force averaged over $4\pi$ sr is stronger than gravity. This beaming could be accomplished by a disk \citep{nakano89a, nakano95a, jijina96a} or an outflow cavity \citep{krumholz05a}, or by any other non-spherical feature that might be found in the flow.

\begin{figure}[t]
\sidecaption
\centerline{\includegraphics[scale=0.25]{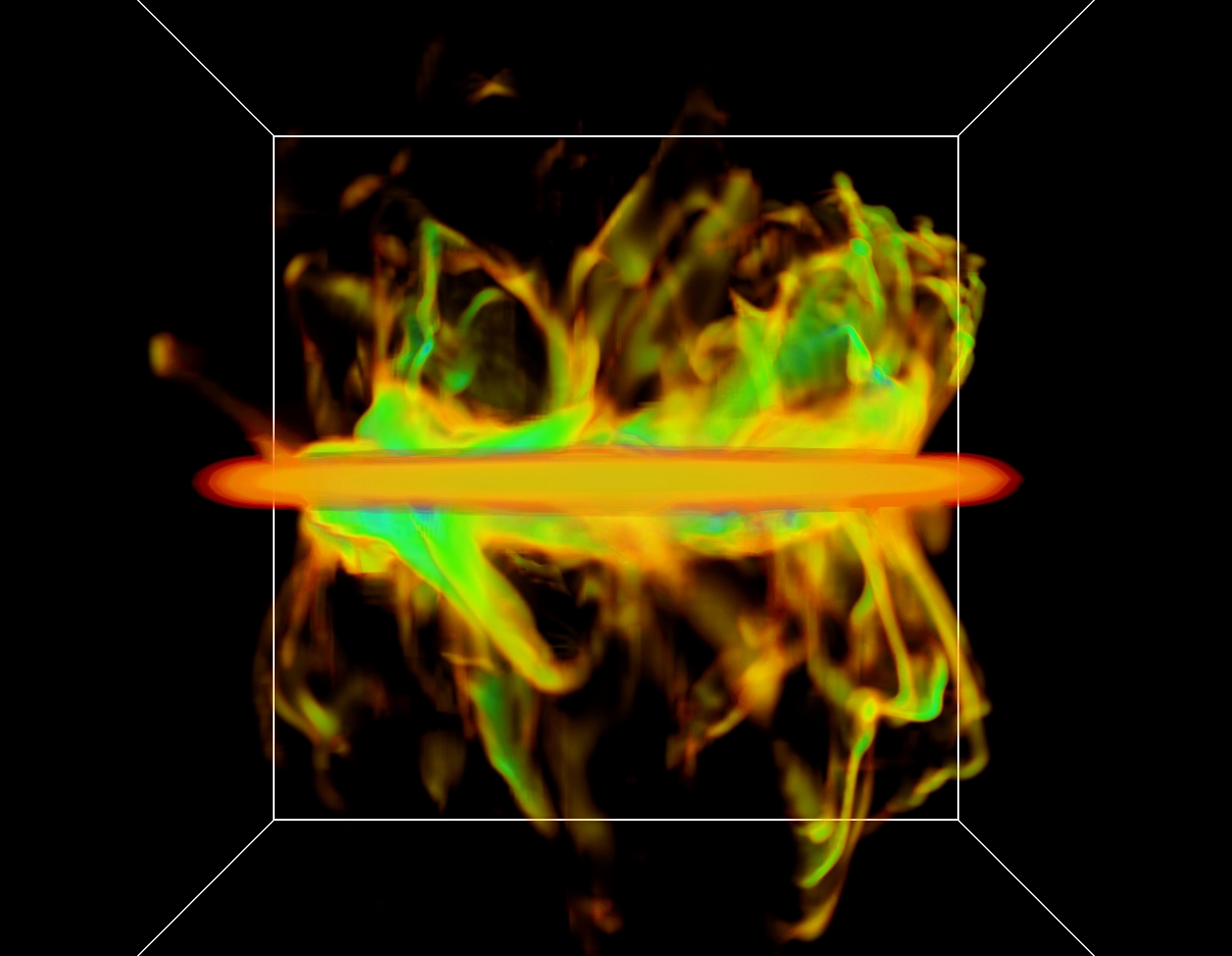}}
\centerline{\includegraphics[scale=0.25]{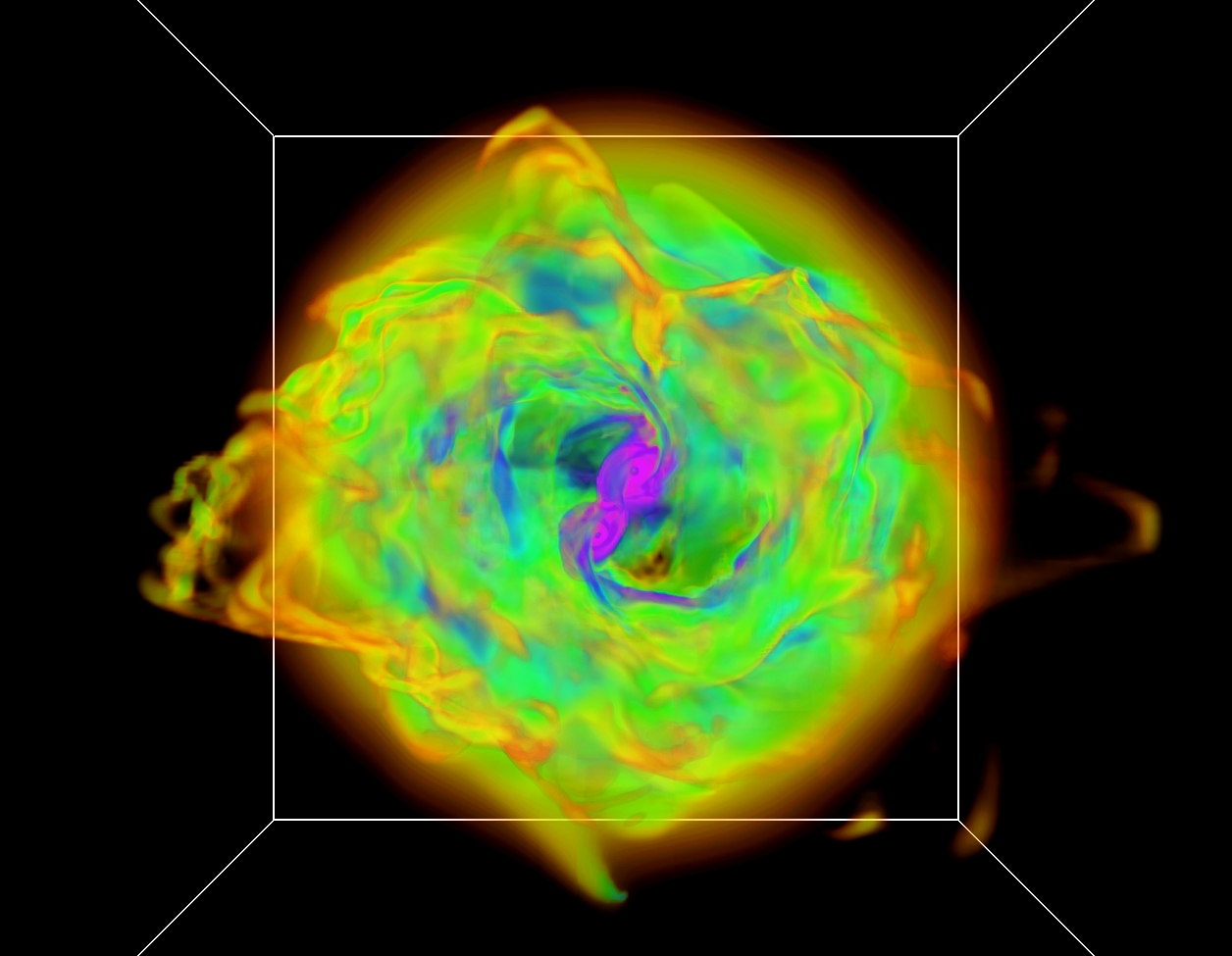}}
\caption{
Volume renderings of the density field in the central 4000 AU of a simulation of the formation of a massive binary system including radiation pressure feedback \citep{krumholz09c}. The top image shows the edge-on view of the disk, while the bottom image shows the face-on view. At the time shown in these images, the simulation contains a $41.5+29.2$ $M_\odot$ binary, each with its own disk, and with the two disks embedded in a circumbinary disk. The filamentary structure above and below the disk is created by radiation Rayleigh-Taylor instability, and consists of dense filaments carrying mass onto the disk.
\label{fig:krumholz09}
}
\end{figure}

The first radiation-hydrodynamic simulations in two dimensions found that beaming by the disk was indeed effective at channeling radiation away from an accreting star \citep{yorke95a, yorke99a, yorke02a}, but that nevertheless the radiation field was able to reverse the accretion flow and prevent formation of stars larger than $\sim 40$ $\msun$. The first three-dimensional radiation-hydrodynamic simulation, on the other hand, found that there was no flow reversal, and that mass was able to accrete essentially without limit \citep{krumholz09c}. Figure \ref{fig:krumholz09} shows a snapshot from this simulation. The key physical process uncovered in these simulations was radiation Rayleigh-Taylor instability (RRTI): the configuration of a radiation field attempting to hold up a dense, accreting gas is unstable to the development of fingers of high optical depth material that channel matter down toward the star, while radiation preferentially escapes through low optical depth chimneys that contain little matter. While the instability was first discovered numerically, subsequently \citet{jacquet11a} and \citet{jiang13a} performed analytic stability calculations that allowed them to derive the linear stability condition and linear growth rate for RRTI.

This picture was somewhat complicated by the work of \citet{kuiper10b, kuiper11a, kuiper12a}, who pointed out that the numerical method used in the \citet{krumholz09c}, while it provided a correct treatment of the dust-reprocessed radiation field, did not properly include the radiation force produced by the direct stellar radiation field. When \citeauthor{kuiper12a}~include this effect, they find that the extra acceleration provided to the circumstellar matter is such that gas tends to be ejected from a protostellar core before the RRTI has time to become non-linear. While there is no reason to doubt that the result is correct in the case of an initially-laminar protostellar core as considered by \citeauthor{kuiper12a}, it is unclear how general this result is, since any pre-existing density structure in the core will ``jump-start" the growth of the instability and allow it to become non-linear in far less time. Such pre-existing density structures are inevitable given the regions of massive star formation are highly turbulent \citep[e.g.][]{shirley03a}, and even in the absence of turbulence, gravitational instabilities in the accretion disk will tend to produce large density contrasts and possibly binary systems \citep{kratter10a}.\footnote{Although \citeauthor{kuiper12a}'s simulations are three-dimensional, they cannot model either turbulence of disk fragmentation, because the numerical method they use for radiation transport can only handle a single star whose location is fixed at the origin of their spherical grid.}

While there is debate about the role of RRTI, there is no debate about whether radiation pressure can actually halt accretion. \citet{kuiper11a} and \citet{kuiper13a} find that, even though radiation pressure is able to eject matter in their simulations, it is unable to eject the accretion disk, and thus that accretion can continue onto stars up to essentially arbitrary masses. Similarly, \citet{cunningham11a}, confirming the hypothesis of \citet{krumholz05a}, show that a protostellar outflow cavity produced by a massive star provides an efficient mechanism for radiation to escape, allowing accretion to continue essentially without any limit due to radiation pressure. Indeed, the presence of an outflow cavity removes the need for RRTI to occur, as it provides a pre-existing low-optical depth chimney through which radiation escapes.  Thus, the consensus of modern radiation-hydrodynamic simulations of massive star formation that radiation pressure does not represent a serious barrier to the formation of stars up to essentially arbitrary masses.

\subsection{Ionization Feedback}

The third potential problem in forming very massive stars is photoionization: galactic molecular clouds generally have escape speeds below 10 km s$^{-1}$ \citep[e.g.,][]{heyer09a}, the sound speed in $\sim 10^4$ K photoionized gas. As a result, if the gas in a star-forming region becomes ionized, the gas pressure may drive a thermal wind that will choke off accretion. This process is thought to be a major factor in limiting the star formation efficiency of giant molecular clouds \citep[e.g.,][]{whitworth79a, matzner02a, krumholz06d}. However, it is much less clear whether photoionization can limit the formation of individual massive stars. The key argument on this point was first made by \citet{walmsley95a}, who noted that an accretion flow onto a massive star can sharply limit the radial extent of an H~\textsc{ii} region. This is simply a matter of the ionizing photon budget: the higher the mass inflow rate, the higher the density of matter around the star, and thus the higher the recombination rate and the smaller the Str\"omgren radius. If the radius of the ionized region is small enough that the escape speed from its outer edge is $>10$ km s$^{-1}$, then photoionized gas will not be able to flow away in a wind or escape. This problem was first considered by \citet{walmsley95a}, who considered an accretion flow in free-fall onto a star, and showed that the escape speed from the edge of the ionized region will exceed the ionized gas sound speed $c_i$ if the accretion rate satisfies
\begin{equation}
\dot{M}_* > \left[\frac{8\pi \mu_{\rm H}^2 G M_* S}{2.2\alpha_B \ln(v_{\rm esc,*}/c_i)}\right] = 4\times 10^{-5} \left(\frac{M_*}{100\,M_\odot}\right)^{1/2} \left(\frac{S}{10^{49}\mbox{ s}^{-1}}\right)^{1/2} \,M_\odot\mbox{ yr}^{-1},
\end{equation}
where $M_*$ is the stellar mass, $S$ is the star's ionizing luminosity (photons per unit time), $\mu_{\rm H}=2.34\times 10^{-24}$ is the mean mass per H nucleus, $\alpha_B\approx 2.6\times 10^{-13}$ cm$^3$ s$^{-1}$ is the case B recombination coefficient, and $v_{\rm esc,*}$ is the escape speed from the stellar surface. The factor of $2.2$ in the denominator arises from the assumption that He is singly ionized. The numerical evaluation uses $v_{\rm esc,*}=1000$ km s$^{-1}$ and $c_i=10$ km s$^{-1}$, but the numerical result is only logarithmically-sensitive to these parameters. Thus an accretion rate of $\sim 10^{-4}$ $\msun$ yr$^{-1}$ is sufficient to allow continuing accretion onto even an early O star. Given the dense, compact environments in which massive stars appear to form, such high accretion rates are entirely expected \citep{mckee03a}. \citet{keto03a} extended \citeauthor{walmsley95a}'s result by deriving a full solution for a spherical inflow plus ionization front in spherical symmetry, and reached the same qualitative conclusion. \citet{keto06a}, \citet{klaassen07a}, and \citet{keto08a} provide direct observational evidence for accretion in photoionized regions.

\begin{figure}[t]
\sidecaption
\centerline{\includegraphics[scale=0.5]{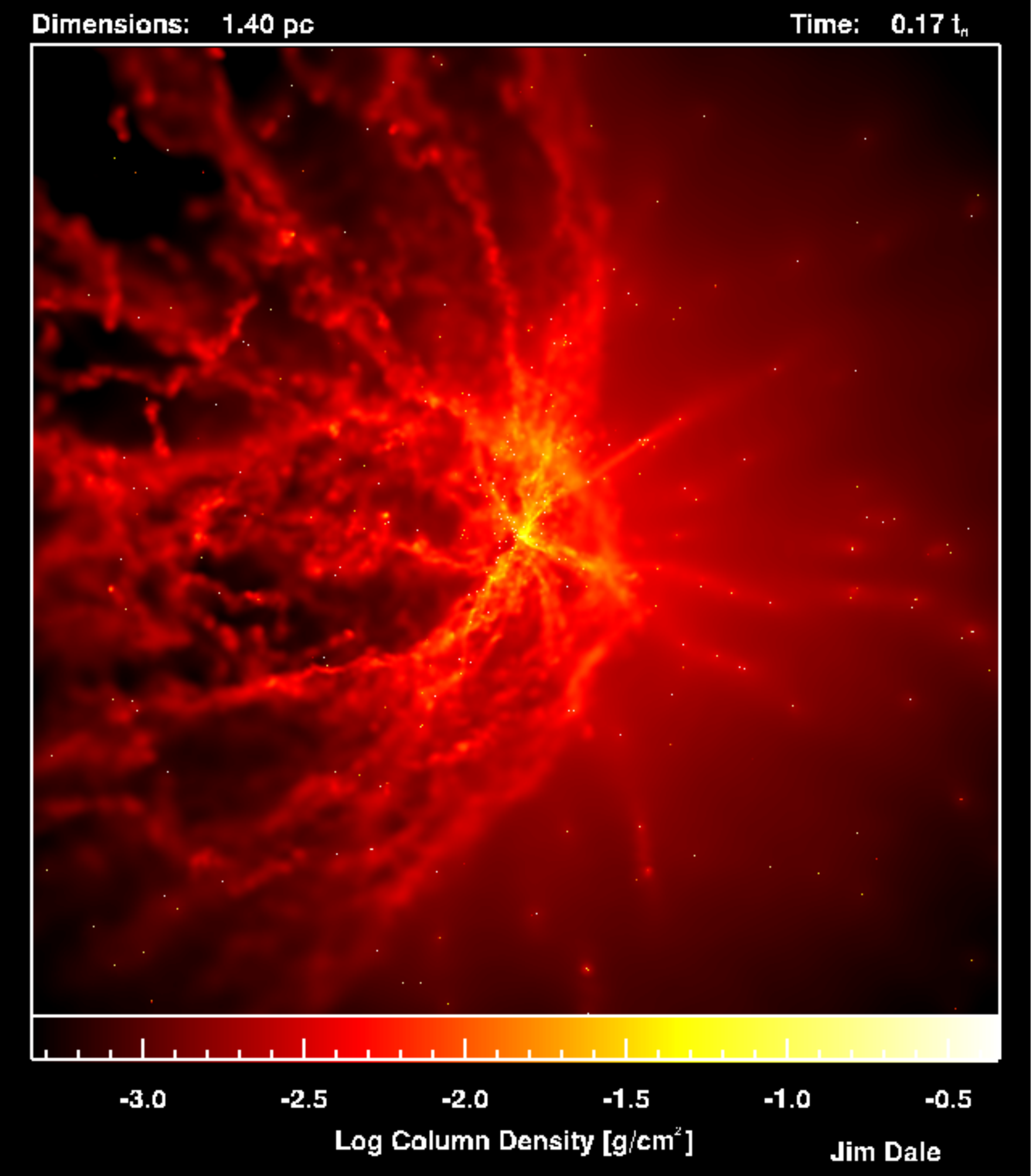}}
\caption{
Column density from a simulation of the formation of a massive star cluster including photoionization feedback \citep{dale05a}. The central star begins the simulation with a mass of $\approx 30$ $M_\odot$, but continues to grow over the course of the simulation, reaching $>100$ $M_\odot$ by the end. White dots are stars.
\label{fig:dale}
}
\end{figure}

This argument makes clear that whether photoionization can limit accretion onto massive stars depends critically on the interplay between the initial conditions, which determine the accretion rate, and stellar evolution, which determines the ionizing luminosity. If the accretion rate drops low enough, and the ionizing flux is high enough, then the ionized region will extend out to the point where photoioinzed gas can escape and accretion will be choked off. The geometry of the flow matters as well. \citet{keto07a} considered rotating infall, and showed that this may result in a configuration where the ionized region blows out in the polar direction, but continues uninhibited through a denser equatorial disk that self-shields against the ionizing photons. In three dimensions turbulent structure may plan an analogous role. \citet{dale05a} and \citet{peters10a, peters11a} have simulated the formation of massive stars and star clusters including photoionization feedback, and they find that photoionization is generally unable to disrupt accretion flows. In the simulations, accretion tends to be highly aspherical, proceeding through disks and filaments, as illustrated in Figure \ref{fig:dale}. Because these structures are dense, they have very high recombination rates and thus are resistant to being photoionized. The structure that tends to result in these simulations is that there are low-density ionized regions where material is escaping, but that the majority of the mass is contained in dense filaments where it continues to accrete. As a result, in \citeauthor{dale05a}'s simulations accretion is able to continue to masses of several hundred $M_\odot$, while in \citeauthor{peters10a}'s simulations reach a mass of $\sim 70$ $M_\odot$ without photoionization halting accretion.

There has been considerably more work on whether ionization can halt accretion in the context of the formation of the first stars. \citet{mckee08a} developed an analytic model for several forms of feedback, and argued that photoionizing radiation will blow out the polar regions of a rotating accretion flow around an accreting star once its mass reaches $\sim 50-100$ $M_\odot$, will thereafter go on to photoevaporate the disk. This process will halt accretion at a mass of $\sim 150$ $M_\odot$. \citet{hosokawa11a} conducted 2D simulations and obtained results qualitatively consistent with \citeauthor{mckee08a}'s model, but with an even lower limiting mass of $\sim 40-50$ $M_\odot$. Similar limiting masses were obtained from three-dimensional simulations by \citet{stacy12a} and \citet{susa13a}, and in a 2D simulation of metal free-star formation in gas that was externally ionized before collapsing (so-called population III.2 star formation), \citet{hosokawa12a} found an even lower limiting mass of 20 $M_\odot$.

The fairly low limiting masses found in the simulations of primordial star formation appear to be in some tension with the results of the numerical simulations of present-day star formation. At first one might think that the presence or absence of dust opacity provides an obvious explanation for the difference, but it is not clear if this is the case. Even at Solar metallicity, most ionizing photons are absorbed by hydrogen atoms and not dust grains (see the Appendix of \citealt{krumholz09d} for a discussion of why this is), so dust is responsible for removing only a small fraction of ionizing photons. Similarly, primordial H~\textsc{ii} regions have somewhat higher temperatures (due to lack of metal lines cooling) and metal-free stars have somewhat higher ionizing luminosities (due to the lack of metal opacity in the stellar atmosphere)

A more promising explanation has to do with the initial conditions, which determine the accretion rate and geometry of the inflow. In an isolated star-forming core, which is the initial condition employed in the primordial calculations, once the photoionized region escapes from the vicinity of the star it can choke off further accretion onto the disk, leaving the disk subject to photoevaporation. However this does not appear to happen in a flow that is continuously fed by large amounts of mass supplied from larger, $\sim 1$ pc scales, as occurs in the present-day star formation simulations. This mass supply into the filaments and disks appears to shield them against photoevaporation. If the initial conditions are the key difference, then for the case of present-day star formation this suggests that the mass limit imposed by photoionization is likely to depend on the large-scale environment, though  exactly which environmental properties are important remains uncertain.

Finally, as a caveat, it is important to note that the treatments of ionizing radiative transfer used in the codes for the simulation of both the present-day and primordial star formation are based on a simple ray-trace using the ``on-the-spot" approximation. In this approximation, one treats ionizing photons produced by recombinations in the ionized gas as having a mean free path of zero, so that photons produced by a recombination to the ground state are re-absorbed on the spot rather than propagating a finite distance. Thus the diffuse radiation field produced by recombinations is ignored, and shadowing is too perfect. This is potentially problematic for the treatment of accretion disks, as the photoevaporation of disks is probably dominated by the diffuse photons produced in the photoionized atmosphere above the disk, rather than by direct stellar radiation \citep{hollenbach94a, mckee08a}. Thus it is unclear if the numerical results are reliable. The question of whether photoionization might limit stellar masses thus remains an only partially-solved problem.

\subsection{Stellar Winds}

A final potential challenge for the formation of massive stars by accretion has received far less theoretical attention: stellar winds. Once the surface temperatures of stars exceed $\sim 2.5\times 10^4$ K, they begin to accelerate fast, radiatively-driven winds \citep{leitherer92a, vink00a, vink01a}. Zero-age main sequence stars reach this temperature at a mass of $\sim 40$ $M_\odot$, and stars of this mass have such Kelvin-Helmholtz timescales that, even if they are rapidly accreting, their radii and surface temperatures during formation are likely to be close to their ZAMS values \citep{hosokawa09a}. The momentum carried by these winds is about half that of the stellar radiation field \citep{kudritzki99a, richer00a, repolust04a}, and so if the direct stellar radiation field cannot stop accretion then the momentum carried by stellar winds will not either.

However, winds might yet be important, because the wind launch velocity is quite large, $\sim 1000$ km s$^{-1}$. As a result, when the winds shock against the dense accretion flow, their post-shock temperature can be $>10^7$ K, well past the peak of the cooling curve \citep{castor75a, weaver77a}, and as a result the gas will stay hot rather than cooling radiatively. Should it become trapped, this hot gas could exert a pressure that is far greater than what would be suggested by its launch momentum -- in effect, it could convert from a momentum-driven flow to an energy-driven one \citep[cf.][]{dekel13a}. If this were to happen, it is possible that the stellar wind gas might be able to interfere with accretion.

There has been a great deal of work on the interaction of post-shock stellar wind gas with the ISM on the scale of star clusters \citep[e.g.,][]{tenorio-tagle07a, dale08a, rogers13a}. This work suggests that the wind gas tends, much like radiation, to leak out through openings in the surrounding dense gas rather than becoming trapped and building up a large pressure. Indeed, on the cluster scale observations appear to confirm that the pressure exerted by the hot gas is subdominant \citep{lopez11a}. However, there is no comparable work on the scale of individual stars, and it is therefore possible that the situation there could be different. Moreover, even when the wind gas does escape on cluster scales, as it flows past the colder, denser material it tends to entrain and carry of some of it. Again, the question of whether this might happen to the accretion flows around individual protostars has yet to be addressed. Given the lack of theoretical or observational attention, the best that can be said for now is that, if the interaction between stellar winds on small scales resembles those seen on larger scales, stellar winds are unlikely to set significant limits on the masses to which stars can grow by accretion.

\section{The Formation of Very Massive Stars by Collision}
\label{sec:collision}

The discussion in the previous section indicates that there is no strong argument against the idea that very massive stars form via the same accretion mechanisms that give rise to stars of lower masses. However, it is also possible for very massive stars to form through an entirely different channel: collisions between lower mass stars. The central challenge for forming massive stars via collisions is the very small cross-sectional area of stars compared to typical interstellar separations, and the relatively short times allowed for collisions by the lifetimes of massive stars. Very massive stars are found routinely in clusters with central densities $\sim 10^4$ pc$^{-3}$ \citep[e.g.][]{hillenbrand98a}, and the highest observed central densities in young clusters are $\sim 10^5$ pc$^{-3}$ \citep[their Figure 9]{portegies-zwart10a}, with the possible exception of R136 \citep{selman13a}. If gravitational focusing is significant in enhancing collision rates (usually the case for young clusters), the mean time between collisions in a cluster consisting of stars of number density $n$ and velocity dispersion $\sigma$, each with mass $m$ and radius $r$, is \citep{binney87a}
\begin{equation}
t_{\rm coll} = 7.1 n_4^{-2} \sigma_1 r_0^{-1} m_0^{-1} \mbox{ Myr},
\end{equation}
where $n_4 = n/10^4$ stars pc$^{-3}$, $\sigma_1 = \sigma/10$ km s$^{-1}$, $r_0 = R/R_\odot$, and $m_0 = m/M_\odot$. Thus under observed cluster conditions, we expect $<1$ collision between 1 $M_\odot$ stars to occur within the $\sim 4$ Myr lifetime of a massive star. Collision rates for stars more massive than the mean require a bit more care to calculate, but even under the most optimistic assumptions, production of very massive stars via collisions requires that clusters reach stellar densities much higher than the $\sim 10^4$ pc$^{-3}$ seen in young clusters. This dense phase must be short-lived, since it is not observed. Models for the production of massive stars via collision therefore consist largely of proposals for how to produce such a short-lived, very dense phase. In this section I examine the collisional formation model. In sections \ref{ssec:gasaccretion} and \ref{ssec:nbody} I describe two possible scenarios by which collisions could occur, and I discuss the mechanics of the collisions themselves, and the role of stellar evolution in mediating collisions, in Section \ref{ssec:stellarevol}.

\subsection{Gas Accretion-Driven Collision Models}
\label{ssec:gasaccretion}
 
The first class of proposed mechanisms to raise the density high enough to allow collisional growth consists of processes that occur during the formation of a star cluster when it is still gas-rich. In a gas-rich cluster, stars can accrete gas, and this process is dissipative: it reduces the total gas plus stellar kinetic energy of the system, with the lost energy going into radiation from accretion shocks on the surfaces of protostars, and from Mach cones created by supersonic motion of stars through the gas. To see why this should lead to an increase in density, it is helpful to invoke the virial theorem. For a system where gravity, thermal pressure, and ram pressure are the only significant forces, the Lagrangian virial theorem states that \citep{chandrasekhar53a, mestel56a}
\begin{equation}
\label{eq:virial}
\frac{1}{2}\ddot{I} = 2\mathcal{T} - \mathcal{W},
\end{equation}
where
\begin{eqnarray}
I & =&  \int r^2\, dm \\
\mathcal{T} & = & \int \left(\frac{3}{2}P + \frac{1}{2}\rho v^2\right)\, dV\\
\mathcal{W} & = & -\int \rho \mathbf{r}\cdot\nabla \Phi\, dV
\end{eqnarray}
are the moment of inertia, the total kinetic plus thermal energy, and the gravitational binding energy, respectively\footnote{The functional form of $\mathcal{W}$ is independent of whether or not there is an external gravitational field, but one can only identify the quantity $\mathcal{W}$ as the gravitational self-energy if the field is entirely due to self-gravity, with no external contribution.}. The quantity $\Phi$ is the gravitational potential. If there are significant forces on the surface of the region, or significant magnetic forces, additional terms will be present, but for the moment we will ignore them.

The process of shock dissipation reduces $\mathcal{T}$ while leaving $\mathcal{W}$ unchanged, so the right-hand side becomes negative, and, on average, the system will tend to accelerate inward to smaller radii. This infall converts gravitational binding energy to kinetic energy, so both $\mathcal{T}$ and $-\mathcal{W}$ rise by the same amount. Because of the factor of $2$ in front of $\mathcal{T}$ in equation (\ref{eq:virial}), this tends to push the right-hand side back toward zero: the system is re-virializing, but at a smaller radius, higher density, and larger kinetic and (in absolute value) binding energy. However, this new equilibrium will last only as long as shocks do not keep decreasing $\mathcal{T}$. If shocks continue to happen, this will drive a continuous decrease in radius, and a continuous rise in density of both gas and stars. \citet{bonnell98a} proposed the first version of this model, and argued that it could drive stellar densities to $\sim 10^8$ pc$^{-3}$, at which point collisions would become common and massive stars could build up in this manner. The required density can be lowered significantly if all massive stars are in primordial hard binaries \citep{bonnell05a}, but even for such a configuration a significant rise in stellar density compared to observed conditions is required. 

\citeauthor{bonnell98a}~suggested that contraction would halt only when gas was removed by feedback from the forming massive stars. This halts contraction because, once the gas is removed, there is no longer a dissipation mechanism available to reduce $\mathcal{T}$. However, \citet{clarke08a} subsequently realized that at sufficiently high density two-body relaxation would become faster than dissipation, and this would halt further shrinkage. In terms of the virial theorem, the increase in $-\mathcal{W}$ required to increase $\mathcal{T}$ and balance the dissipation starts to come from stars forming tight binaries rather than from overall shrinkage of the system. The maximum density that can be reached therefore depends on the total cluster mass, in such a manner as to prevent collisions from becoming significant in clusters substantially smaller than $\sim 10^4$ $M_\odot$. It is important to note that this excludes the Orion Nebula Cluster, which contains a star of $\approx 38$ $M_\odot$ \citep{kraus09b}, suggesting that stars of at least this mass at least can form via non-collisional processes.

\begin{figure}[t]
\sidecaption
\centerline{\includegraphics[scale=1]{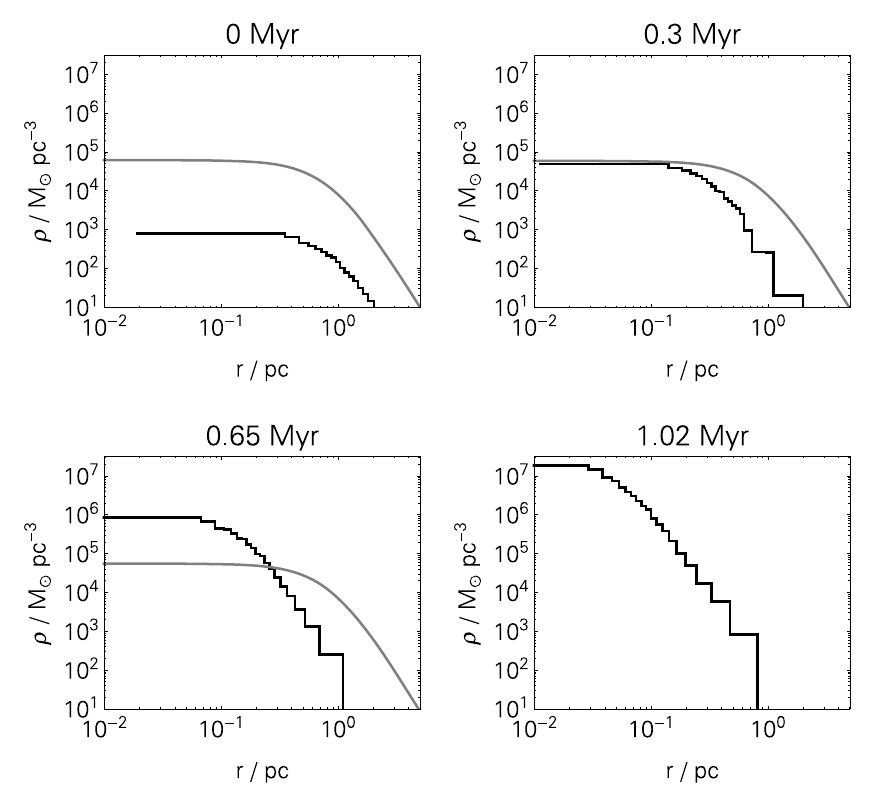}}
\centerline{\includegraphics[scale=1.5]{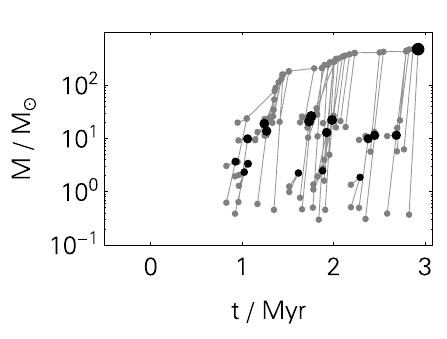}}
\caption{
Example results from the N-body plus accretion simulations of \citet{moeckel11a}. The top set of four panels shows the radially-averaged stellar density profile as a function of time in the simulations (black lines), together with the mass profile corresponding to the imposed gas potential (gray lines). The bottom panel shows the growth history of some of the most massive stars in the simulations. Points indicated stellar masses and the times when those stars first appear, and the convergence of two lines indicates a merger between two stars that yields a more massive star. Black points indicate stars that survive to the end of the simulation, while gray points indicate stars that merge before the end of the simulation.
\label{fig:moeckel}
}
\end{figure}

These conclusions were based on analytic models, but more recently \citet{moeckel11a} and \citet{baumgardt11a} conducted N-body simulations including analytic prescriptions for the effects of gas accretion.\footnote{These authors did not include the effect of gas drag due to Mach cones, which for Bondi-Hoyle accretion flows is actually a factor of several larger than the change in stellar momentum due to accretion \citep{ruffert94b}, but this is probably not the most serious limitation of the simulations.} In these models, the gas is treated as a fixed potential that is reduced as the stars gain mass, eventually disappearing entirely when a prescribed amount has been accreted; this sets the limit on the duration of the gas-dominated phase. Figure \ref{fig:moeckel} shows an example output from one of these simulations. As anticipated by \citet{clarke08a}, in these models the stars sink to the center until they form a stellar-dominated region in which two-body relaxation inhibits further contraction, though these regions can also undergo core collapse (see the next section). They both find that, as a result of this limitation, stellar collisions during the gas-dominated phase are negligible unless the initial conditions are already very compact or massive, with half-mass radii of $\sim 0.1$ pc or less and/or masses of $\sim 10^4$ $M_\odot$ or more.

The requirement for very high initial densities creates significant tension with observations. \citet{moeckel11a} find that even the Arches cluster is insufficiently dense to have produced stellar collisions up to this point, despite the fact that it contains numerous very massive stars. Similarly, \citet{baumgardt11a} find that, once the gas potential is removed and clusters re-virialize, those that began their evolution at densities high enough to induce significant numbers of collisions end up far too compact in comparison to observed open clusters, including those containing very massive stars. As a result of these findings, both sets of authors tentatively conclude that stellar collisions during the gas-dominated phase cannot be the dominant route to the formation of very high mass stars, though they cannot rule out the possibility that such collisions occur in rare circumstances.

Before relying on these conclusions too heavily, it is important to understand the limitations of these calculations. Undoubtedly the largest one is the simple prescription used to treat the gas. In these models, the shape of the gas potential (though not its depth) is fixed, the accretion rates onto stars are fixed and independent of stellar mass or position, and the final star formation efficiency is also fixed. Obviously none of these assumptions are fully realistic. In particular, depending on the effectiveness of stellar feedback, the gas potential might either shrink and promote increases in stellar density, or the gas potential might vary violently in time as gas is pushed around by stellar feedback, pumping energy into the stars and preventing contraction -- indeed, the latter is seen to occur in at least some simulations that do treat the gas \citep{li06b, nakamura07a, wang10a}. It is unclear how the conclusions might change if these phenomena were treated more realistically.

\subsection{Gas-Free Collision Models}
\label{ssec:nbody}
 
The second class of models for inducing growth of very massive stars via collisions takes place in the context of gas-free clusters. These mechanisms, and their potential role in young massive clusters, were recently reviewed by \citet{portegies-zwart10a}, and I refer readers there for further details. The advantage of this approach compared to the gas-driven one is that the time available for collisions is longer, but the disadvantage is the lack of gas drag as a mechanism for raising the density.

Clusters of equal-mass stars are unstable to spontaneous segregation into a contracting core and an expanding envelope, in which the negative heat capacity of the system drives a continuous transfer of energy out of the core and thus ever-higher densities \citep{lynden-bell68a}. In a cluster containing a spectrum of masses, contraction of the core is further enhanced by the tendency of the stars to mass-segregate, with the core consisting of more massive, dynamically-cool stars, and the envelope consisting of low-mass, dynamically-hot ones \citep{spitzer69a, gurkan04a}. While there is no doubt that these processes operate, it is uncertain whether they are fast enough to produce collisions within the $\sim 4$ Myr lifetime of the most massive stars. \citet{portegies-zwart99a} conclude based on N-body simulations that collisions will occur before massive stars die enough if the central density starts at $\sim 10^7$ stars pc$^{-3}$. In this case, the mergers themselves are dissipative and will trigger further core contraction, leading to runaway formation of a single very massive object. As in the case of gas-driven collisions, the existence of a large population of primordial hard binaries can somewhat reduce the density required to initiate this cascade \citep{portegies-zwart02a}. Even so, the initial densities required in the simplest gas-free collision models would preclude the possibility of very massive stars forming by collisions except perhaps in R136. Models in which a significant fraction of very massive stars form via collision therefore generally posit a set of initial conditions that significantly increases the collision rate.

One way to shorten the time required for core collapse and the onset of collisions is to consider a cluster with primordial mass segregation, meaning that the cluster is mass-segregated even before the gas-free evolution begins \citep[e.g.][]{ardi08a, goswami12a, banerjee12a, banerjee12b}. Such a starting configuration reduces the time requires for core collapse to begin because it provides both a higher density and a higher mean stellar mass (and thus a lower relaxation time) in the cluster center. Depending on the degree of mass segregation, the reduction in time to the onset of core collapse and collisions can be $\sim 1-2$ Myr, a non-trivial fraction of the lifetime of a very massive star, and simulations using sufficiently mass-segregated initial conditions generally find that collisions become common before massive stars end their lives.

The extent to which star clusters actually are primordially mass-segregated is unclear. Observations generally show at least some degree of mass segregation in present-day clusters, but the amount varies widely. At the low-mass end of clusters containing massive stars, in the Orion Nebula Cluster the Trapezium stars are all at the cluster center, but there is no observed mass segregation for any stars except these \citep{hillenbrand98a, huff06a}. In NGC 3603 \citep{pang13a} and R136 \citep{andersen09a}, the cluster is segregated throughout so that the mass function is flatter at small radii, but more massive stars are more segregated than less massive ones. However, all of these clusters are $\sim 1-3$ Myr old, so it is entirely possible that the segregation we see now is a product of dynamical evolution during this time, not primordial segregation -- indeed, \citet{pang13a} argue that the segregation they observe in NGC 3603 is more consistent with dynamical evolution from a weakly-segregated or unsegregated initial state than with primordial segregation. Unfortunately answering this question fully would require that observations probe the gas-enshrouded phase, which is possible only in the infrared, where low resolution creates severe difficulties with confusion. Indeed, confusion is a serious worry for measurements of mass segregation even in optically-revealed clusters \citep{ascenso09a}.

Another way to accelerate the dynamical evolution of star clusters is to begin from unrelaxed initial conditions. Both observations \citep{furesz08a, tobin09a} and simulations \citep{offner09b} of star clusters that are still gas-embedded show that the stars are subvirial with respect to the gas, and such cold conditions lead to more rapid dynamical evolution than virialized initial conditions \citep{allison09a}. Larger star clusters may also be assembled via the mergers of several smaller clusters within the potential provided by a massive gas cloud \citep{bonnell03a, mcmillan07a, fujii12a}. These smaller clusters, since they have smaller numbers of stars, also have smaller time-scales for core collapse. \citet{allison09a} find that substructured initial conditions accelerate mass segregation, but it is unclear whether they do so enough to accelerate collisions. \citet{fujii13a} find that the extent to which the formation of a large cluster out of multiple star clusters influences collisions depends on the ratio of the assembly time to the core collapse time of the initial subclusters. If the subclusters undergo core collapse before merging, then they may have a few internal collisions, but there are no collisions in the merged cluster, and collisional growth of stars is negligible overall. On the other hand, if core collapse does not occur in the subclusters before they merge, the evolution is similar to that of a cluster that formed as a single entity.

In summary, collisions during the gas-free phase are unlikely to contribute significantly to the growth of very massive stars if star clusters are born virialized and non-segregated, but in reality neither of these assumptions is likely to be exactly true. The viability of collisional formation models then turns sensitively on the extent to which these assumptions are violated, and this question is unfortunately poorly constrained by observations. Hydrodynamic simulations of the formation of massive star clusters that include the gas-dominated phase may be helpful in addressing this question, but to be credible these simulations will need to include feedback processes such as stellar winds, photoionization, and radiation pressure that are presently omitted from most models.

\subsection{Stellar Evolution and Massive Star Mergers}
\label{ssec:stellarevol}

One important subtlety for models of the growth of massive stars via mergers is that the outcome depends not just on N-body processes, but also on the physics of stellar collisions, and on the structure and subsequent evolution of stellar merger products. Both questions have been the subject of considerable study in the context of mergers between low-mass stars leading to the production of blue stragglers, but only a few authors have conducted similar simulations for mergers involving massive stars. Mergers involving massive stars (particularly very massive ones) may be substantially different than those involving low-mass stars because of the importance of radiation pressure for massive stars. As stellar mass  increases, the increasing dominance of radiation pressure brings the structure close to that of an $n=3$ polytrope, which is very weakly bound. Moreover, radiative forces may be non-negligible during the collision itself. For example, just as radiation pressure may be capable of inhibiting accretion, it may be capable of ejecting material that is flung off stellar surfaces during a collision, thereby increasing mass loss during the collision.

One quantity of interest from stellar merger simulations is the amount of prompt mass loss during the collision itself. Models for collisional growth generally assume that the mass loss is negligible, thus maximizing the collisional growth rate. \citet{freitag05a}, \citet{suzuki07a}, and \citet{glebbeek13a} all find in their simulations that losses are indeed small, with at most $\sim 10\%$ of the initial mass being ejected for realistic collision velocities. In three-body mergers produced when an intruder enters a tight binary system, the loss can be as large as $\sim 25\%$ \citep{gaburov10a}. However, a very important limitation of these simulations is that they do not include any radiative transfer, and treat radiation pressure as simply an extra term in the equation of state, with the radiation pressure determined by the matter temperature, which in turn is determined by hydrodynamic considerations. This is likely to be a very poor approximation for the material that is flung outward from a collision, where illumination from the central merged object will dominate the thermodynamics, as it does in the case of accretion onto massive stars. In the approximation used by the existing merger simulations, it is impossible for radiation pressure to eject matter, and thus the $\sim 5-10\%$ mass loss found in the simulations simply represents the mass of material that is raised to escape velocities during the collision itself. This figure should therefore be thought of as representing a lower limit. There is a clear need to reinvestigate this problem using a true radiation-hydrodynamics code. If the mass loss has been underestimated, collisional growth will be harder than is currently supposed.

A second quantity of interest from merger simulations is the radius of the merger product. When stars merge, shocks during the collision raise the entropy of the stellar material, so that when hydrostatic equilibrium is re-established a few days after the merger, the resulting star will initially be very bloated compared to main sequence stars of the same mass, and will gradually shrink over a Kelvin-Helmholtz timescale \citep{dale06a, suzuki07a}. Building up very massive stars via collisions likely requires multiple mergers, and at the very high densities required, the interval between mergers may be smaller than the KH timescale, so that the growing stars will have enlarged radii. Whether this will enhance or reduce the rate of collisional growth is unclear. \citet{suzuki07a} point out that the enhanced radii of the merger products make them bigger targets that are more likely to collide with other stars. On the other hand, \citet{dale06a} note that the envelopes of the post-merger stars are even more weakly bound than those of massive main sequence stars, and as a result such collisions may actually erode the envelope rather than add to it, ultimately limiting collisional growth. Which effect dominates is unclear, as no merger simulations involving such swollen stars have been reported in the literature.

\begin{figure}[t]
\sidecaption
\centerline{\includegraphics[scale=1]{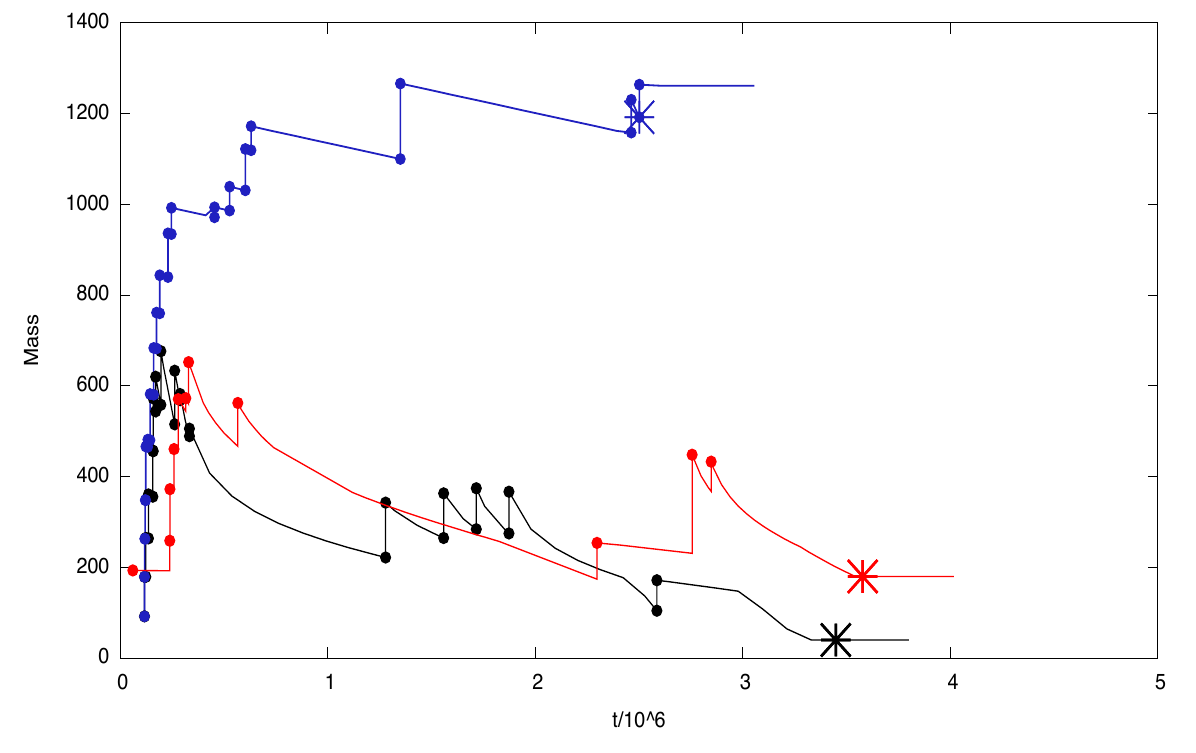}}
\caption{
Results from three simulations of massive stellar mergers driven by gas-free core collapse by \citet{vanbeveren09a}, reprinted by permission. The blue line shows a calculation using fairly modest wind mass losses, similar to those adopted by \citet{portegies-zwart99a}. The black line shows a calculation with identical initial conditions but using a wind prescription taken from \citet{belkus07a} for Solar metallicity stars, and the red line is the same but using a metallicity of 5\% of Solar.
\label{fig:vanbeveren}
}
\end{figure}

A final consideration for collisional growth models in the gas-free phase, where the timescales involved may be several Myr, is mass loss via stellar winds. At masses below $\sim 100$ $M_\odot$, wind mass loss rates are considerable, but are unlikely to be able to counteract the effects of collisional growth if the density is high enough for runaway merging to begin. However, little is known about wind mass loss rates at still higher masses, and there are good empirical arguments that they might be orders of magnitude larger \citep{belkus07a}. N-body simulations using these enhanced winds find that they remove mass from stars faster than collisions can add it, yielding only very transient growth to large masses, followed by rapid shrinkage back to $\sim 100$ $M_\odot$ \citep{yungelson08a, vanbeveren09a, glebbeek09a}. Figure \ref{fig:vanbeveren} provides an example. Moreover, the winds might remove mass so efficiently that they reduce the gravitational potential energy of the system fast enough to offset the loss of kinetic energy that occurs during mergers, halting the collisional cascade completely.

\section{Observational Consequences and Tests}
\label{sec:discrimination}

Having reviewed the various models for the origins of very massive stars, I now turn to the question of their predictions for observable quantities, and how these might be used to test the models. One can roughly divide these predictions into those that apply on the scale of star clusters, and those that apply on the scale of individual star systems.

\subsection{The Shape of the Stellar Mass Function}

On the cluster scale, one obvious difference between collisional and accretion-based models of massive star formation is their predictions for the form of the stellar mass function at the massive end -- note that I refer here to the present-day mass function (PDMF) rather than the initial mass function (IMF), since in gas-free collision models very massive stars are absent in the IMF and only appear later due to collisions. As discussed above, in the case where massive stars form by the same accretion processes that produce low-mass stars, one naturally expects that very massive stars should simply be a smooth continuation of the Salpeter mass function seen at lower masses. The situation is very different for collisional models, where very massive stars form via a fundamentally different process than low mass ones. Not surprisingly, this different formation mechanism gives rise to a feature in the stellar mass function.

\begin{figure}[t]
\sidecaption
\centerline{\includegraphics[scale=1]{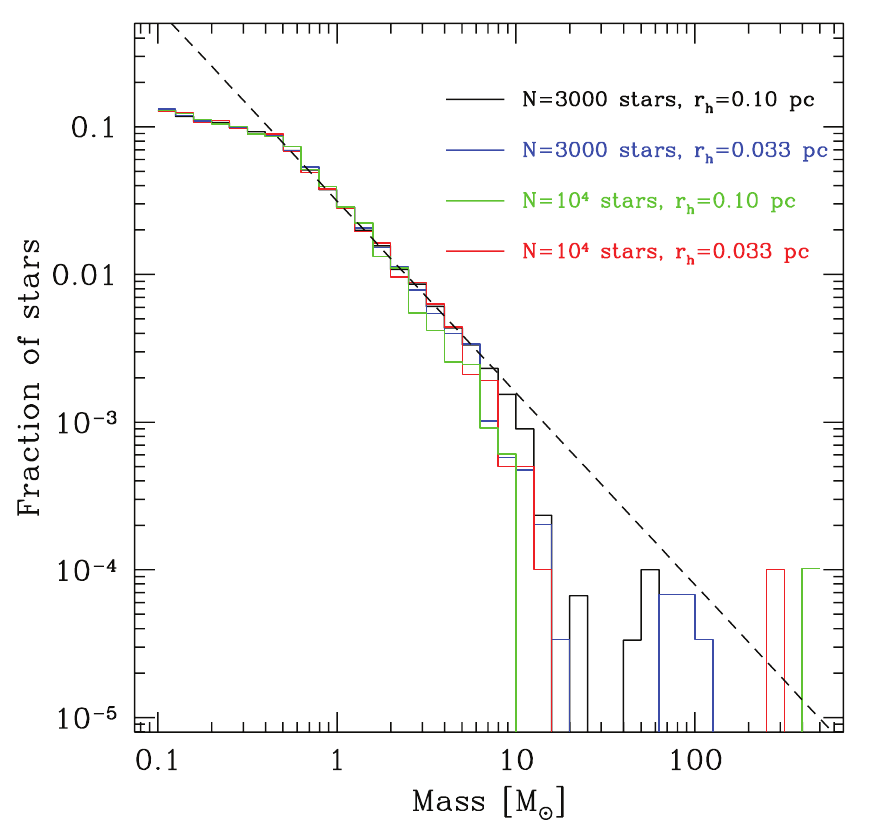}}
\caption{
Mass functions of stars in an N-body simulation of gas-driven stellar collisions by \citet{baumgardt11a}. The histograms are mass functions obtained 10 Myr after the beginning of the simulation, and the number of stars and initial half-mass radius used in each of the simulations are as indicated in the legend. The straight dashed line is the Salpeter mass function. Note that the simulations all predict a turndown in the mass function relative to Salpeter at masses from $\sim 10-100$ $M_\odot$.
\label{fig:baumgardt11}
}
\end{figure}

For the gas-driven collision gas, both \citet{moeckel11a} and \citet{baumgardt11a} find that the typical outcome of collisions is one or two objects whose masses are much greater than those of any other object, and a corresponding depletion of objects slightly less massive than the dominant one or two. Figure \ref{fig:baumgardt11} shows an example result from \citet{baumgardt11a}. As the plot shows, collisions that yield very massive stars of several hundred $M_\odot$ tend to produce an overall mass function in which the range from $\sim 10-100$ $M_\odot$ is actually significantly under-populated relative to the Salpeter slope found at lower masses, while the one or two most massive objects that are formed by collision represent a significant over-population relative to Salpeter. Unfortunately the authors of models in which collisional growth occurs during the gas-free phase have generally not reported the full mass functions produced in their simulations, but given that the mechanism for assembling the very massive stars is essentially the same as in the gas-driven models -- runaway collisions that agglomerate many massive stars into one or two supermassive ones -- it seems likely that these models would predict a similar functional form for the mass function.

At present there is no observational evidence for mass functions of this form. \citet{massey98b} report that the mass function in R136 is well-approximated by a single powerlaw with a Salpeter-like slope from $3-120$ $M_\odot$, and \citet{andersen09a} report a continuous powerlaw slope over an even wider range of mass, with no evidence for a turn-down in the vicinity of 10 $M_\odot$.  Similarly, \citet{espinoza09a} examine the Arches cluster and report that the mass function above 10 $M_\odot$ is well-described by a powerlaw with a slope $\Gamma=-1.1\pm 0.2$, consistent within the errors with the Salpeter value $\Gamma = -1.35$. There are significant systematic uncertainties on these values, arising mostly from the challenge of assigning masses to stars based on photometry, but it is important to note that a mass function of the form predicted the collisional simulations should be apparent even from the \textit{luminosity} function, independent of the mapping between luminosity and mass. Due to confusion, even luminosity functions can be difficult to measure in the cores of clusters dense enough to be candidates for collisions, but this data should improve significantly in the era of 30m-class telescopes. Observations with these facilities should be able to settle the question of whether the mass and luminosity functions in cluster cores show the characteristic signature of a depletion from $\sim 10-100$ $M_\odot$ coupled to a one or two stars at a few hundred $M_\odot$ that is predicted by collisional models.

\subsection{Environmental-Dependence of the Stellar Mass Function}

A second possible discriminant between accretion and collision as mechanisms for the formation of very massive stars is the way the stellar mass function depends on the large-scale properties of the cluster. As noted above, both gas-free and gas-driven collision models require very high stellar densities (even in the present-day state) and very high cluster masses; \citet{baumgardt11a} argue that clusters where gas-driven collisions occur all end up too compact compared to observed ones, and \citet{moeckel11a} argue that the Arches is not dense enough to be able to produce significant collisions. In contrast, accretion models either predict that the stellar IMF will be independent of environment, or that massive stars will be biased to regions of high surface density \citep{krumholz08a, krumholz10a}. Accretion models do not predict that there should be an upper limit to stellar masses that is a function of either cluster mass or central stellar density.

This is a somewhat weaker test than the functional form of the stellar mass function, simply because the model predictions are somewhat less clear, but it may nonetheless provide a valuable complement. The challenge here will be obtaining a sample large enough to see if there is a statistically-significant correlation between the presence of absence of stars above some mass and properties of the environment like cluster mass or density. The major challenge is that one expects a correlation between maximum stellar mass and cluster size simply due to size of sample effects. Observations must therefore remove the size of sample effect statistically, and search for a small correlation that might remain once the size of sample effect is removed. Some authors have claimed to detect such a correlation already in Galactic clusters \citep{weidner10a, weidner13a}, while others have reported the absence of any such correlation in extra-Galactic environments \citep{calzetti10b, fumagalli11a, andrews13a}. Given the poorly-understood selection issues associated with the Galactic sample (which is culled from the literature, rather than produced by a single survey), it seems likely that the extragalactic results based on uniform samples are more reliable, but the issue remains disputed.

An observation of a very massive star formed in relative isolation would also be strong proof that collisions are not required to make such stars, though it would not rule out the possibility that some stars form that way. There are several candidates for isolated stars with masses above $\sim 20$ $M_\odot$ \citep[and in some cases as much as $100$ $M_\odot$;][]{bressert12a, oey13a}, and which appear unlikely to be runaways because they have small radial velocities and no bow shocks indicating large transverse motions. However, there remains the possibility that these are ``slow runaways" with that were ejected very early and thus managed to reach fairly large distances from the cluster despite their fairly small space velocities \citep{banerjee12a}.

\subsection{Companions to Massive Stars}

The properties of massive star companions provide a final potential test for distinguishing accretion-based and collisional formation models. It is well known that massive stars are much more likely that stars of lower mass to be members of multiple systems. \citet{mason09a} report a companion fraction of $75\%$ for O star primaries in Milky Way star clusters\footnote{O stars outside clusters are likely to have been dynamically ejected from the cluster where they were born, and in the process stripped of companions.}, while \citet{sana13a} find a companion fraction of 50\% for O stars in the Tarantula Nebula in the Large Magellanic Cloud. \citet{sana12a} estimate that 70\% of O stars have a companion close enough that they will exchange mass with it at some point during their main sequence or post-main sequence evolution, and that $1/3$ of O stars have a companion so close that they will merge\footnote{Mergers and mass transfer may also be significant during pre-main sequence evolution -- see \citet{krumholz07c}.}. From the standpoint of formation theories, a high binary fraction is expected regardless of whether massive stars are formed via accretion \citep[e.g.][]{kratter08a, kratter10a, krumholz12b} or collisions \citep[e.g.][]{portegies-zwart99a, bonnell05a}. However, much less is known about the prevalence of low-mass companions to massive stars, or to tight massive binaries, and the statistics of low-mass companions to massive stars provide another potential test of formation models.

Accretion-based models predict that, in addition to their massive companions, massive stars are also very likely to have low-mass companions at separations of $\sim 100-1000$ AU \citep{kratter06a, kratter08a, kratter10a, krumholz12b}. The authors of collisional models have not thus far published detailed predictions for massive binary properties, but these should be trivial to obtain from the simulations already run, and it seems likely that the dense dynamical environment required for collisions would strip any low-mass, distant companions from massive stars. Thus observations capable of probing large mass ratios at intermediate separations might be a valuable test of massive star formation models.

\begin{figure}[t]
\sidecaption
\centerline{\includegraphics[scale=1]{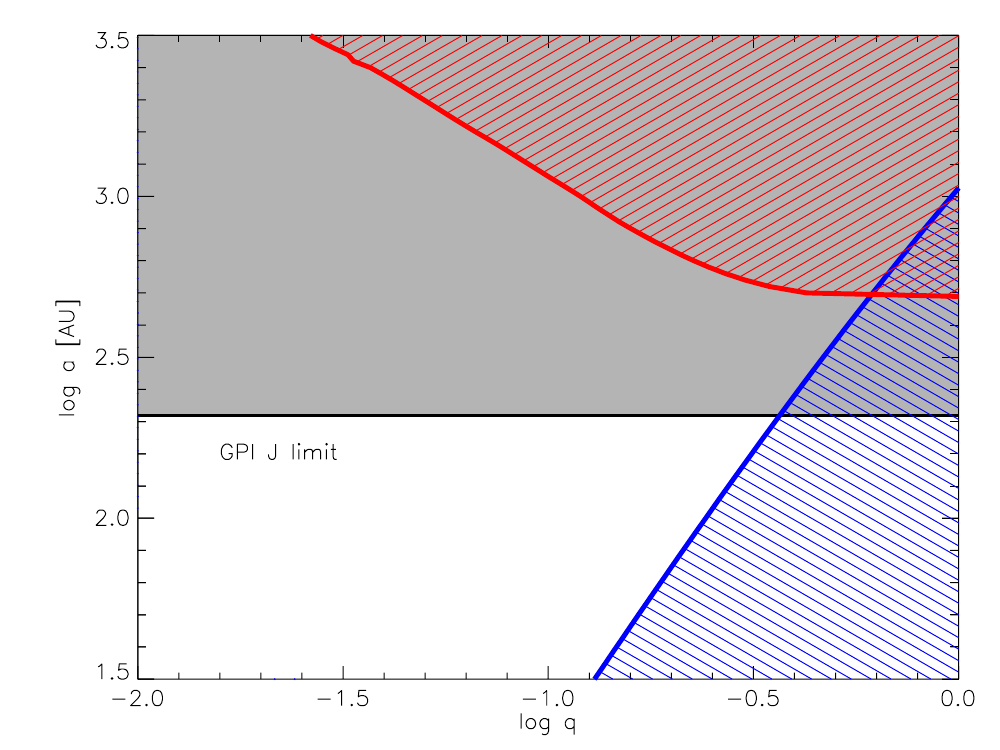}}
\caption{
Estimated detectability of companions to massive stars as a function of mass ratio $q$ and semi-major axis $a$ using spectroscopic surveys (blue hashed region), adaptive optics imaging surveys (red dashed region), and using a next-generation instrument like GPI (gray region). These sensitivities are calculated for a hypothetical primary of mass $M_p = 60$ $M_\odot$ at a distance $d=2$ kpc. The limit for spectroscopy is computed using equation (\ref{eq:aspec}) assuming a velocity semi-amplitude limit $v_{\rm lim} = 5$ km s$^{-1}$. The limit for adaptive optics is computed from equation (\ref{eq:aao}) using the Padova mass-magnitude relations for ZAMS stars \citep{marigo08a, girardi10a}. The GPI limit shown is the physical size corresponding to the $0\farcs{11}$ size of the GPI occulting stop in J band.
\label{fig:binary_limits}
}
\end{figure}

This range is unfortunately relatively hard to probe via observations, as the expected radial velocity shifts are too small to be measured against the broad lines of a massive star, and the large contrast ratio makes direct imaging difficult. Observations using high contrast techniques like speckle imaging \citep{mason09a}, adaptive optics \citep{sana10a}, and lucky imaging \citep{maiz-apellaniz10a} are starting to push into this range, but still have some distance to go. Consider a primary massive star of mass $M_p$ with a companion of mass $q M_p$ (with $q<1$) in a circular orbit with semi-major axis $a$. The system is a distance $d$ from the Sun. Spectroscopic surveys are generally limited in their ability to detect companions by the velocity semi-amplitude $v_{\rm lim}$ to which they are sensitive, which is generally $\sim 5-10$ km s$^{-1}$ depending on the linewidths of the primary star \citep[e.g.][]{kiminki07a, kobulnicky07a}. The companion is detectable only if
\begin{equation}
\label{eq:aspec}
a < \left(\frac{q^2}{q+1}\right)\frac{G M_p}{v_{\rm lim}^2} \approx 5.3 \left(\frac{q}{0.1}\right)^2 \left(\frac{M_p}{60\,M_\odot}\right)\left(\frac{v_{\rm lim}}{10\mbox{ km s}^{-1}}\right)^{-2}\mbox{ AU}.
\end{equation}
Imaging surveys are limited by the contrast they can achieve. For example, \citet{sana10a} estimate that their detection threshold is a contrast of $\Delta K_s \approx \Delta K_{s,0} (r-0\farcs{24})^{1/3}$, where $\Delta K_{s,0} = 6$ mag and $r$ is the angular separation in arcsec and $\Delta K_s$ is the contrast in the $K_s$ band. Given a mass-magnitude relationship $K_s(M)$, a companion will be detectable if
\begin{equation}
\label{eq:aao}
|K_s(M_p) - K_s(q M_p)| < \Delta K_{s,0} \left[2.06\times 10^5\left(\frac{\mbox{arcsec}}{\mbox{rad}}\right)\frac{a}{d}-0\farcs{24}\right]^{1/3}.
\end{equation}
Figure \ref{fig:binary_limits} shows the ranges of $q$ and $a$ over which companions to massive stars are detectable given these sensitivities.

The next generation of high-contrast systems designed for planet imaging, such as the Gemini Planet Imager (GPI) and Spectro-Polarimetric High-contrast Exoplanet Research instrument (SPHERE) should push much farther and be able to detect even very low mass companions to massive stars. Indeed, the contrast ratios these instruments can achieve is such that, outside of their occulting stops, they should be sensitive to companions to O stars down to the hydrogen burning limit. Figure \ref{fig:binary_limits} shows an estimate of the sensitivity region for GPI. Observations using these instruments provide a definitive census of massive star companions at high mass ratio and intermediate separation. This is likely to prove a powerful constraint on formation models.

\section{Conclusions and Summary: Does Star Formation Have an Upper Mass Limit?}
\label{sec:masslimit}

Having discussed the two main formation scenarios, I now return to the question of whether star formation has a mass limit. To review, there is at present no really convincing evidence that any mechanism is capable of halting the growth of stars by accretion. The classical mechanism for limiting stellar masses is radiation pressure, but non-spherical accretion, produced by some combination of disks, outflow cavities, and instabilities appears to defeat this limit. Similarly, the problem of gas fragmenting too strongly to form massive stars appears to be solved by a combination of radiative heating and magnetic fields, though the possibility that disk fragmentation might at some point limit stellar masses remains. Photoionization and stellar winds are somewhat more promising as mechanisms that might limit stars' growth, but these remain at best possibilities. There are no real analytic models applicable to present-day (as opposed to primordial) star formation that suggest what limits these mechanisms might impose, and there are no simulations demonstrating that either of these processes are capable of terminating accretion. A fair description of the state of the field a decade ago might have been that the presumption was in favor of feedback limiting massive star formation, and that the burden of proof was on those trying to show that feedback could be overcome. The last decade of work has reversed that situation, with all tests thus-far performed showing that accretion is very difficult to reverse. This does not prove that no mechanism can limit stellar masses, but does mean that such a limit would need to be demonstrated.

For collisions, the question is not whether but where they can create very massive stars. There is no doubt that collisions and collisional growth will occur if the conditions are dense enough, and the only question is the frequency with which such dense conditions are created in nature, which in turn will determine the contribution of the collisional formation channel to the overall population of very massive stars. No presently-observed star cluster has a density high enough for collisions to be likely, but it is possible that these clusters experienced a very dense phase during which collisions occurred. This could have been either an early gas-dominated phase or a later phase of core collapse aided by primordial mass segregation and high levels of primordial substructure. However, the threshold density required to achieve significant collisional growth depends on details of massive star mergers and winds that are poorly understood. Even for favorable assumptions about these uncertain parameters, it is not clear that the observed present-day properties of massive star clusters can be reconciled with an evolutionary history in which they were once dense enough to have produced collisions.

There are a number of observational tests that may be able to settle the question of which of these mechanisms is the dominant route to the formation of the most massive stars. Accretion models predict that massive stars are simply the tip of the iceberg of normal star formation, so that the high end of the stellar mass function is continuous and does not depend radically on the environment, and that massive stars are likely to have low-mass as well as high-mass companions. Although the observable consequences of the collisional formation models have received somewhat less attention, such models appear to predict quite different results: there should be a large gap in stellar mass functions separating the bulk of the accretion-formed stellar population from the few collisionally-formed stars, and this feature should appear only in the most massive and densest clusters. It seems likely that these collisionally-formed stars will lack low-mass companions. It should be possible to perform most or all of these observational tests with the coming generation of telescopes and instruments, which will provide higher angular resolution and contrast sensitivity than have previously been possible.

\begin{acknowledgement}

I thank all the authors who provided figures for this review: H.~Baumgardt, J.~Dale, N.~Moeckel, A.~C.~Myers, and D.~Vanbeveren. During the writing of this review I was supported by NSF CAREER grant AST-0955300, NASA ATP grant NNX13AB84G, and NASA TCAN grant NNX14AB52G. I also thank the Aspen Center for Physics, which is supported by NSF Grant PHY-1066293, for hospitality during the writing of this review.

\end{acknowledgement}

\bibliographystyle{apj.bst}
\bibliography{refs}

\end{document}